\documentclass[doublecol]{epl2}
\usepackage{amsmath}
\usepackage{graphicx}
\usepackage{subeqnarray}

\title{Persistence in Random Walk in Composite Media}

\author{D.Chakraborty \footnote{email: tpdc2@mahendra.iacs.res.in}}
\institute{ Department of Theoretical Physics,\\
Indian Association for the Cultivation of Science,\\
Jadavpur, Kolkata-700032, India.}


\abstract{
We consider a class of inhomogeneous media known as composite media that is often encountered in experimental sciences and investigate the persistence probability of a random walker in such a system. Analytical and numerical results for the crossover time scales has been obtained for a composite system with two homogeneous components and three homogeneous components respectively. }

\begin{document}

\maketitle

The phenomenon of persistence in various stochastic processes has been well documented over the past decade\cite{1}-\cite{13}.
Even the most simple of all stochastic processes, a random walker in a homogeneous and infinite media, exhibits the phenomenon of persistence and the non trivial persistence exponent $\theta$ has the value $1/2$ \cite{9}. In an experimental setup finite boundaries become important and we have recently investigated the effect of finite boundaries on the survival probability of a random walker in an homogeneous system\cite{16}. Its then a natural question to ask as to how the survival probability behaves for a random walker in an heterogeneous system. Although random walk in spatially disordered media has been already studied \cite{17}-\cite{19} and in few cases exact results are known for a similar quantity ,the first passage time\cite{20}-\cite{23}, little is known about the persistence probability. We shall consider one such class of heterogeneous system which is  known as composite media and is often encountered in experimental science. A composite system essentially comprises of segments of different homogeneous media which differ in their macroscopic properties, such as diffusion coefficients. Redner has already investigated the first passage properties for a diffusive process in such a composite system\cite{24}. He considers a linear chain of $N$ blocks each of length $L_i$ having diffusivities $D_i$. The mean first passage time in such a system is calculated to be 
\begin{equation}
\label{1}
\langle t \rangle =\frac{1}{2} \sum_{i=1}^{N} \frac{L_i^2}{D_i} + \sum_{i<j}^{N} \frac{L_i L_j}{D_j}.	
\end{equation}

While the first passage probability is simply the probability that the particle escapes from one of the boundaries, the persistence probability is different and is defined as the probability that the random walker has not crossed the origin up to time $t$. If the system was homogeneous then the persistence probability of a random walker would be simply $p(t) \sim t^{-\theta}$ with $\theta=1/2$. A composite system is slightly different in the sense that near the boundaries the difference in diffusivities tend to give a bias to random walker. 

The simplest of composite media that can be constructed is the one with two homogeneous segments with different diffusivities. We shall first derive the result for such a system and later generalize this result for different types of composite media.

Consider two homogeneous media of diffusivities $D_1$, henceforth called medium 1 and $D_2$, medium 2. A slab of medium 1 is placed between $-L$ and $+L$ and the rest of the space is filled with medium 2 as shown in Fig. 1. Since the diffusion coefficients are different, it follows that the stochastic noise correlator will be different and in particular they are
\begin{equation}
\label{2}
\langle \eta_i(t_1) \eta_j(t_2) \rangle =2D_i \delta_{ij} \delta(t_1-t_2),	
\end{equation}
where $i,j$ are the medium indices $1$ and $2$ and $t_1>t_2$.
For a random walker the the probability that the walker is at $(x,t)$ starting from $(x_0,0)$ simply obeys the diffusion equation in two segments as
\begin{subeqnarray}
\label{3}
\slabel{3a}
\frac{\partial P}{\partial t}=D_1 \frac{\partial^2 P}{\partial x^2} \phantom{1cm} -L\le x \ge L\\ 
\slabel{3b}	
\frac{\partial P}{\partial t}=D_2 \frac{\partial^2 P}{\partial x^2} \phantom{1cm} -L > x > L
\end{subeqnarray}

The exact dynamics of the problem can solved by considering the Laplace transform of Eq.(\ref{3a}) and Eq.(\ref{3b}) in which case the solution to the equations become 
\begin{subeqnarray}
\label{4}
\slabel{4a}
P(x,s)=A_1 \exp\left(-\sqrt{\frac{s}{D_1}}|x|\right) \\
\nonumber
+A_2 \exp\left(\sqrt{\frac{s}{D_1}}|x|\right) \quad -L \le x \ge L \\
\slabel{4b}
P(x,s)=A_3 \exp\left(-\sqrt{\frac{s}{D_2}}|x|\right) 
\end{subeqnarray}

The coefficients $A_1, A_2, A_3$ are found from the boundary conditions that the probability $P(x,t)$ and the current $-D\frac{\partial P}{\partial x}$ is continuous across the boundary. Finally, the third unknown coefficient is found from the normalization of the probability. The resulting expression, however, is complicated and it is difficult to extract any information from it. 

\begin{figure}
\onefigure[scale=0.3]{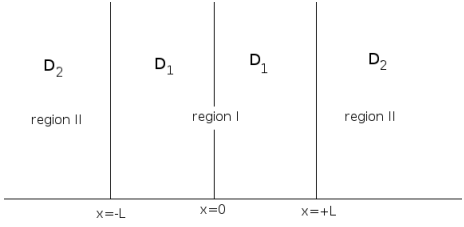}
\caption{Arrangement of two homogeneous media.}
\end{figure}

We, instead, take a different approach to derive our result. The equation of motion for the random walker is not changed in spite of the heterogeneity of the system and is simply
\begin{subeqnarray}
\label{5}
%\slabel{5a} 
\frac{\mathrm{d} x}{\mathrm{d} t} = \eta_1(t) \qquad \textrm{for}
\quad -L \le x \le L, \\
%\slabel{5b)
\frac{\mathrm{d} x}{\mathrm{d} t} = \eta_2(t) \qquad \textrm{for}
\quad -L > x > L.
\end{subeqnarray}

Let the time required for a random walker to reach the boundary $x=\pm L$ be $\tau$. In which case we can write down the solution for the equation of motion in the two different regions. For $-L<x<L$ the solution is simply
\begin{equation}
\label{6}
	x(t)=\int_{0}^{t} \mathrm{d} t' \eta_1(t'),
\end{equation}
whereas for $-L>x>L$, when the particle is in medium 2 the solution for $x(t)$ becomes
\begin{equation}
\label{7}
	x(t)= \int_{0}^{\tau} \mathrm{d} t' \eta_1(t') +\int_{\tau}^{t} 
\mathrm{d} t' \eta_2(t').
\end{equation}
Eq.(\ref{7}) simply states the fact the random walker has spent time $\tau$ in medium 1 and the rest of the time in medium 2. Both Eq.(\ref{7}) and Eq.(\ref{8}) are valid when the walker is deep inside medium 1 or medium 2, since none of the equations considers the hopping across the boundary. When deep inside either media the multiple boundary crossings are rare events whereas when the walker is near the boundary multiple crossings are frequent and it is due to these multiple crossing events the crossover is not sharp and there will be two crossover time scales in the problem.

The correlation $\langle x(t_1) x(t_2) \rangle$ can now be worked out carefully. First consider the case $-L<x(t_1)<L$, $-L<x(t_2)<L$ and $t_1>t_2$. In this case the correlator becomes
\begin{equation}
\label{8}
\langle x(t_1) x(t_2) \rangle=\int_{0}^{t_1}\mathrm{d}t'_1\int_{0}^{t_2}
\mathrm{d}t'_2 \quad \langle \eta_1(t'_1) \eta(t'_2) \rangle=2D_1 t_2\\
\end{equation}
where we have used Eq.(\ref{2}) for the noise correlator. If, however, $-L>x(t_1)>L$, $-L<x(t_2)<L$ and $t_1>t_2$ then we have
\begin{eqnarray}
\label{9}
\nonumber
\langle x(t_1) x(t_2) \rangle = \langle\left[\int_{0}^{t_2} \mathrm{d} t'_2 \eta_1(t'_2) \right]\times \\
\left[ \int_{0}^{\tau} \mathrm{d} t'_1 \eta_1(t'_1) +\int_{\tau}^{t} \mathrm{d} t'_1 \eta_2(t'_1) \right]\rangle
\end{eqnarray}
Since $\langle \eta_1(t)\eta_2(t) \rangle=0$, the above expression simplifies to
\begin{equation}
\label{10}
\langle x(t_1) x(t_2) \rangle =\int_{0}^{\tau}\mathrm{d} t'_1 \int_{0}^{t_2} \mathrm{d} t'_2 \langle \eta_1(t'_2)  \eta_1(t'_1) \rangle
\end{equation}
As $\tau >t_2$ the correlation $\langle x(t_1) x(t_2) \rangle$ becomes
\begin{equation}
\label{11}
\langle x(t_1) x(t_2) \rangle = 2D_1 t_2 
\end{equation}
Finally, for $-L>x(t_1)>L$ and $-L>x(t_2)>L$, $t_1>t_2 $, the correlator becomes
\begin{eqnarray}
\label{12}
\nonumber
\langle x(t_1) x(t_2) \rangle =\langle\left[\int_{0}^{\tau} \mathrm{d} t'_1 \eta_1(t'_1) +\int_{\tau}^{t} \mathrm{d} t'_1 \eta_2(t'_1) \right]\times\\
\left[ \int_{0}^{\tau} \mathrm{d} t'_2 \eta_1(t'_2) +\int_{\tau}^{t} \mathrm{d} t'_2 \eta_2(t'_2)\right]\rangle
\end{eqnarray}
Since the cross-correlation of the noise is zero we arrive at
\begin{eqnarray}
\label{13}
\nonumber
\langle x(t_1) x(t_2) \rangle = \int_{0}^{\tau} \mathrm{d}t'_1 \int_{0}^{\tau} \mathrm{d}t'_2 \langle \eta_1(t'_1) \eta_1(t'_2) \rangle\\
+ \int_{\tau}^{t} \mathrm{d} t'_1 \int_{\tau}^{t} \mathrm{d} t'_2 \langle\eta_2(t'_1) \eta_2(t'_2)\rangle
\end{eqnarray}
The first term in Eq.(\ref{13}) is easy to calculate and gives us
\begin{equation}
\label{14}
\int_{0}^{\tau} \mathrm{d}t'_1 \int_{0}^{\tau} \mathrm{d}t'_2 \langle \eta_1(t'_1) \eta_1(t'_2)=2D_1\tau 
\end{equation}
To evaluate the second term we make a transformation of variable $t''=t'-\tau$ and we have
\begin{equation}
\label{15}
\int_{0}^{t_1-\tau} \mathrm{d} t''_1 \int_{0}^{t_2-\tau} \mathrm{d} t''_2 2D_2 \delta(t''_1-t''_2) = 2D_2 (t_2-\tau)
\end{equation}
Hence, the complete correlator is 
\begin{equation}
\label{16}
\langle x(t_1)x(t_2)\rangle = 2(D_1-D_2) \tau+2D_2t_2
\end{equation}

Of all the quantities in Eq.(\ref{16}) the only unknown is $\tau$. Since Eq.(\ref{16}) gives us noise averaged quantities we might as well replace $\tau$ by an average value, which is simply $\tau=\frac{L^2}{2D_1}$, the average time for a random walker to reach $x=\pm L$.

It is clear from Eq.(\ref{11}) and Eq.(\ref{16}) that there are two relevant time scales in the problem. The first one is $\tau_1=\left(\frac{D_1-D_2}{D_1}\right)\frac{L^2}{2D_1}$ whereas the second one is $\tau_2=\left(\frac{D_1-D_2}{D_1}\right)\frac{L^2}{2D_2}$. It is between these two time scales when the random walker undertakes multiple hoppings across the boundary, as a result of which the temporal regime $\tau_1<t<\tau_2$ gives the crossover region in the system.

The complete correlator now takes the form
\begin{eqnarray}
\label{17}
\nonumber
\langle x(t_1)x(t_2)\rangle &=& 2 D_1 t_2 \phantom{2 cm} \textrm{for}\quad t_2<\tau_1 ,\\
\nonumber
&=& \left(\frac{D_1-D_2}{D_1}\right)L^2+2D_2 t_2 \quad \textrm{for}\quad 
t_2>\tau_2,\\
\end{eqnarray}

In the limit $D_1=D_2=D$ we get the correct result for a homogeneous medium.

\begin{figure}
\onefigure[scale=0.8]{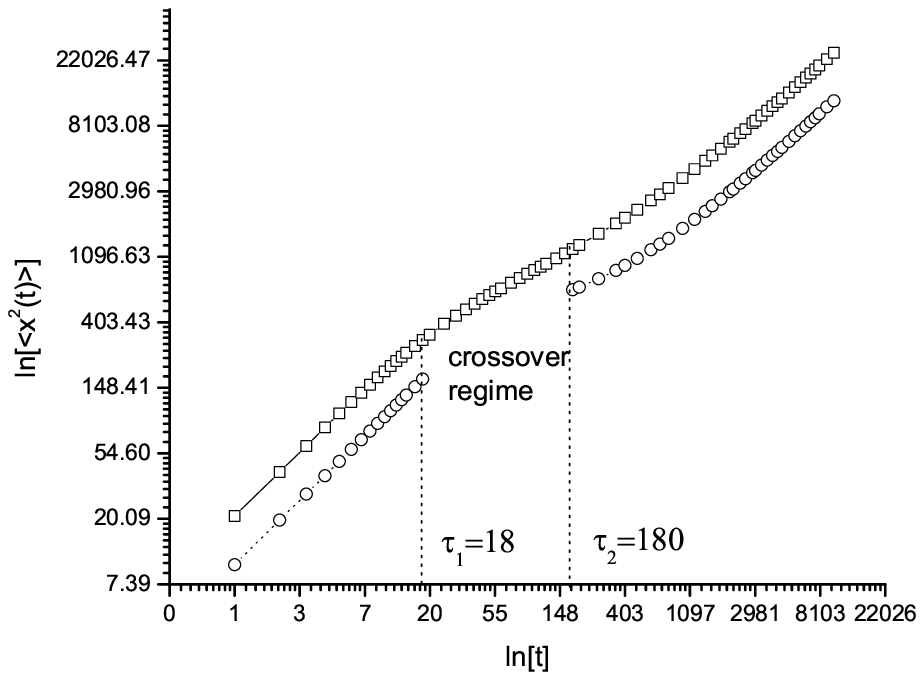}
\caption{Plot of $\langle x^2(t) \rangle$ vs time in log-log scale for
$L=20$, $2D_1=20$, $2D_2=2$. The crossover time scales and the crossover regime is indicated in the figure. The square points are actual data from numerical simulation and the circular points are fit of Eq.(\ref{17})}
\end{figure}

\begin{figure}
\onefigure[scale=0.8]{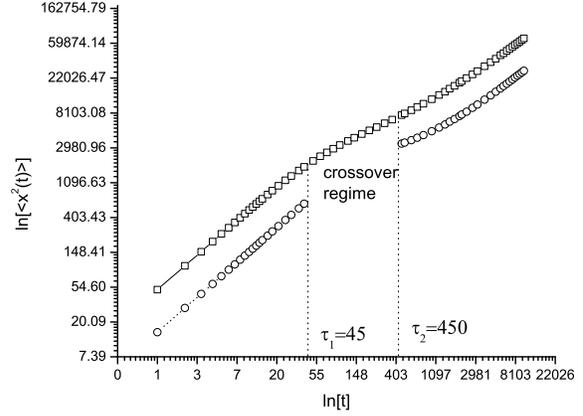}
\caption{Plot of $\langle x^2(t) \rangle$ vs time in log-log scale for 
$L=50$, $2D_1=50$, $2D_2=5$. The crossover time scales and the crossover regime is indicated in the figure. The square points are actual data from numerical simulation and the circular points are fit of Eq.(\ref{17})}
\end{figure}

A numerical simulation of the system for two different values of $L$ and two different sets of $D_1$ and $D_2$ has been done. Simulation result for the mean square displacement $\langle x^2(t)\rangle$ is shown in Fig.2 and Fig.3. Configuration averaging of $10^4$ has been done for both the systems to obtain the result.

To calculate the survival probability in the two regimes we use Eq.(\ref{17}) and with suitable transformations both in $x$ and $t$ convert the process to a Gaussian stationary process.
Define a normalized variable $\bar{X}(t)=x(t)/\sqrt{\langle x^2(t) \rangle}$. The correlator in this normalized variable, $\langle \bar{X}(t_1)\bar{X}(t_2)\rangle$, is then given by
\begin{eqnarray}
\label{18}
\nonumber
\langle \bar{X}(t_1)\bar{X}(t_2)\rangle	&=& \sqrt{\frac{t_2}{t_1}} \quad \textrm{for}\quad t_2<\tau_1 \\ 
\nonumber
&=&\sqrt{\frac{\beta L^2+2D_2 t_2}{\beta L^2+2D_2 t_1}}\quad \textrm{for}\quad t_2>\tau_2 ,\\
\end{eqnarray}
with $\beta=\frac{D_1-D_2}{D_1}$.
For $t<\tau_1$ we define the usual transformation in time, $T=\ln t$ and the two time correlation function in the new time variable becomes
\begin{equation}
\label{19}
\langle \bar{X}(T_1) \bar{X}(T_2)\rangle = e^{-1/2(T_1-T_2)},
\end{equation}
and the survival probability $p(t)$ for this temporal regime, in real time, is then
\begin{equation}
\label{20}
	p(t)\sim t^{-1/2}.
\end{equation}

For $t>\tau_2$ we define a new time variable $T$ as
\begin{equation}
\label{21}
e^{T}=\beta L^2+2D_2 t.
\end{equation}
The correlation function $\langle \bar{X}(T_1) \bar{X}(T_2) \rangle$ takes the form of Eq(\ref{19}), except that the time transformations are different. Since the process is a Gaussian stationary process and the correlator is exponentially decaying, the survival probability in the new time variable, $P(T)$, is
\begin{equation}
\label{22}
	P(T)=e^{-T/2}.
\end{equation}
 In real time the survival probability $p(t)$ takes the form
\begin{equation}
\label{23}
	p(t)\sim \frac{1}{\sqrt{\beta L^2+2D_2 t}}.
\end{equation}

A plot of the survival probability, Eq.(\ref{20}) and Eq.(\ref{23}) for the two time regimes is shown in Fig. 4 and Fig. 5. The crossover timescales $\tau_1$, $\tau_2$ and the crossover regimes are also indicated in the figures. 
Configuration averaging of $10^6$ has been done to obtain the numerical results of Fig. 4 and Fig 5. Theoretical and numerical values of $\tau_1$ and $\tau_2$ are presented in Table A for two different set of values of $L$, $D_1$ and $D_2$. 
\begin{figure}
\onefigure[scale=0.8]{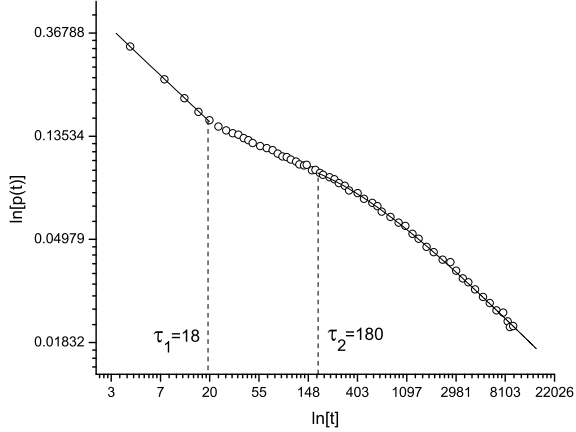}
\caption{Plot of survival probability $ p(t) $ vs time in log-log scale for 
$L=20$, $2D_1=20$, $2D_2=2$. The crossover time scales and the crossover regime is indicated in the figure. The circular points are actual data from numerical simulation and the solid lines are fit of Eq.(\ref{21}) and Eq.(\ref{24})}
\end{figure}

\begin{figure}
\onefigure[scale=0.8]{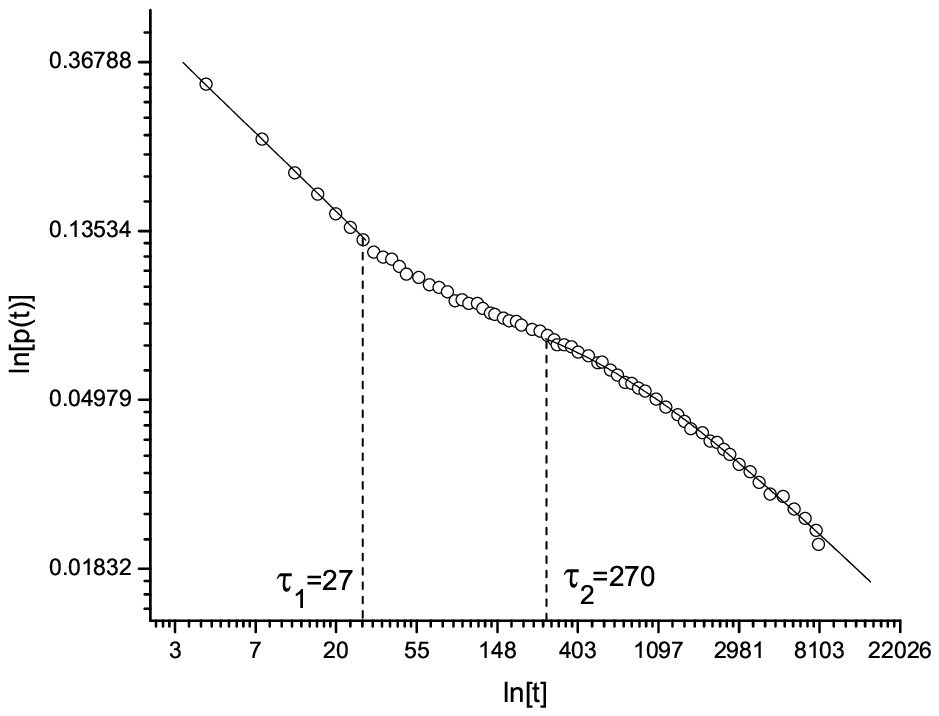}
\caption{Plot of survival probability $p(t)$ vs time in log-log scale for 
$L=30$, $2D_1=30$, $2D_2=3$. The crossover time scales and the crossover regime is indicated in the figure. The circular points are actual data from numerical simulation and the solid lines are fit of Eq.(\ref{21}) and Eq.(\ref{24})}
\end{figure}

\begin{center}
\textbf{Table A}
\end{center}

\begin{tabular}{|c|c|c|c|c|}
	\hline
Parameter Values & $\tau_1^{th}$ & $\tau_1^{nu}$ & $\tau_2^{th}$ & $\tau_2^{nu}$ \\
\hline
$L=20$& & & & \\
$2D_1=20$& 18& 17.161& 180 & 179.988\\
$2D_2=2.0$ & & & & \\
\hline
$L=30$ & & & & \\
$2D_1=30$ & 27& 27.228& 270& 271.899\\ 
$2D_2=3.0$ & & & & \\
\hline
\end{tabular}
\\
\begin{center}
\textbf{\begin{large}Survival probability for three media.\end{large}}
\end{center}

In this section we consider a composite system that is made of three homogeneous media with diffusivities $D_1$, $D_2$ and $D_3$. The medium with diffusivity $D_1$ is placed between $\pm L_1$ while the second medium 
 with diffusivity $D_2$ is placed symmetrically between $-(L_2+L_1)<x<-L_1$ and 
$L_1<x<(L_1+L_2)$ as shown in the figure.
\begin{figure}
\onefigure[scale=0.3]{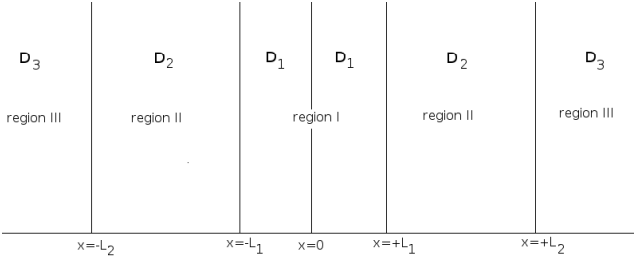}
\caption{Arrangement of three homogeneous media.}
\end{figure}
For a random walker in region I the average time to reach the boundary $\pm L_1$ is $\tau=\frac{L_1^2}{2D_1}$. When the random walker is in region II the average time to cross a region the length $L_2$ is once again $\tau'=\frac{L_2^2}{2D_2}$. Thus, for $t<\tau$ the particle spends its time in region I, for $\tau<t<\tau'$ the walker is in region II while for $t>\tau'$ the walker escapes to region III.
The equation of motion in all the three region are 
\begin{equation}
\label{25}
\frac{\mathrm{d}x}{\mathrm{d}t}=\eta_i(t) \quad \textrm{with i=1,2,3 for three media,}
\end{equation}
with the noise correlator 
\begin{equation}
\label{26}
\langle \eta_i(t) \eta_j(t') \rangle =2 D_i \delta_{ij} \delta(t-t'),
\end{equation}
where $i,j$ are the medium indices running from $1$ to $3$.

The solutions to Eq(\ref{25}) for the three regions are respectively
\begin{equation}
\label{27}
x(t)=\int_0^t \mathrm{d}t'\eta_1(t')\quad \textrm{for $-L_1<x(t)<L_1$}\\
\end{equation}

\begin{equation}
\label{28}
x(t)=\int_0^{\tau} \mathrm{d}t'\eta_1(t')
 +\int_{\tau}^t \mathrm{d}t'\eta_2(t')
 \end{equation}
for $L_1<x(t)<L_2$ and $-L_2<x(t)<-L_1$,

\begin{eqnarray}
\label{29}
\nonumber
x(t)=\int_0^{\tau} \mathrm{d}t'\eta_1(t')
 +\int_{\tau}^{\tau+\tau'}\mathrm{d}t'\eta_2(t')+\int_{\tau+\tau'}^t \mathrm{d} t' 
\eta_3(t')\\
\nonumber 
\textrm{for $x(t)>L_2$ and $x(t)<-L_2$}.\\
\end{eqnarray}

The two time correlation function $\langle x(t_1) x(t_2) \rangle$ can be worked out carefully and for both $x(t_1)$ and $x(t_2)$ lying in region I, with $t_1>t_2$ is 
\begin{equation}
\label{30}
\langle x(t_1) x(t_2) \rangle =2 D_1 t_2
\end{equation}

For $x(t_1)$ and $x(t_2)$ lying in region II, $\langle x(t_1) x(t_2) \rangle$ takes the form
\begin{equation}
\label{31}
\langle x(t_1) x(t_2) \rangle =\beta L_1^2 +2D_2 t_2
\end{equation}

while for $x(t_1)$ and $x(t_1)$ lying in region III, using the fact that the cross correlations of the noise is zero, the correlator becomes
\begin{eqnarray}
\label{32}
\nonumber
\langle x(t_1) x(t_2) \rangle=\int_{0}^{\tau} \mathrm{d}t'_1 \int_{0}^{\tau} \mathrm{d}t'_2 \langle \eta_1(t'_1) \eta_1(t'_2) \rangle\\
\nonumber
+ \int_{\tau}^{\tau+\tau'} \mathrm{d} t'_1 \int_{\tau}^{\tau+\tau'} \mathrm{d} t'_2 \langle\eta_2(t'_1) \eta_2(t'_2)\rangle \\
+\int_{\tau+\tau'}^{t_1} \mathrm{d} t'_1 \int_{\tau+\tau'}^{t_2} \mathrm{d} t'_2 \langle\eta_3(t'_1) \eta_3(t'_2)\rangle.
\end{eqnarray}

The first integral is simply $2D_1\tau$.
The second integral is performed by making use of the transformation $t''=t'-\tau$ and the integral reduces to 
\begin{equation}
\nonumber
\int_{0}^{\tau'}\mathrm{d} t'_1 \int_{0}^{\tau'} \mathrm{d} t'_2 \langle\eta_2(t'_1) \eta_2(t'_2)\rangle =2D_2\tau'
\end{equation}

while for the third integral we use the transformation $t''=t'-(\tau+\tau')$ and the integral is evaluated to be $2D_3 (t_2-\tau-\tau')$.
Hence the correlator becomes
\begin{eqnarray}
\label{33}
\nonumber
\langle x(t_1) x(t_2) \rangle&=&2D_1 \tau +2D_2 \tau' +2D_3 (t_2-\tau-\tau')\\
\nonumber
&=& \beta_1 L_1^2+\beta_2 L_2^2 +2D_3 t_2 ,\\
\end{eqnarray}
with $\beta_1=(D_1-D_3)/D_1$ and $\beta_2=(D_2-D_3)/D_2$.
It is clear from Eq.(\ref{30}), Eq.(\ref{31}) and Eq.(\ref{33}) that there four relevant time scales in the problem. The first one is obviously $\tau_1=\beta \frac{L_1^2}{2D_1}$. The second one is $\tau_2=\beta \frac{L_1^2}{2D_2}$. The temporal regime $\tau_1<t<\tau_2$ represents the crossover regime from region I to region II, when the walker feels the effect of the inhomogeneity. Similarly, the third time scale is $\tau_3=\frac{1}{2D_2}(\beta_1 L_1^2+\beta_2 L_2^2)$ and the fourth time scale is $\tau_4=\frac{1}{2D_3}(\beta_1 L_1^2+\beta_2 L_2^2)$.  $\tau_3<t<\tau_4$ is the crossover regime from region II to region III and it is during this time when the walker spends most of its time near the boundary of region II and region III.  Thus the proper time scales for which Eq.(\ref{30}), Eq.(\ref{31}) and Eq.(\ref{33}) are valid are respectively $0<t<\tau_1$, $\tau_2<t<\tau_3$ and $t>\tau_4$ while the time intervals $\tau_1<t<\tau_2$ and $\tau_3<t<\tau_4$ represents the two crossover regimes.

The mean square displacement $\langle x^2(t) \rangle$ is then
\begin{eqnarray}
\label{34}
\nonumber
\langle x^2(t) \rangle&=&2D_1t \quad\textrm{for $0<t<\tau_1$}\\
\nonumber
&=& \beta L_1^2 + 2D_2 t_2 \quad \textrm{for $\tau_2<t<\tau_3$}\\
\nonumber
&=& \beta_1 L_1^2+\beta_2 L_2^2+2D_3 t \\
\quad \textrm{for $t>\tau_4$}.
\end{eqnarray}

Note that for $D_2=D_3$ we recover the first case, that is a composite media with two homogeneous components while for $D_1=D_2=D_3=D$ we recover the case for a homogeneous system.

To obtain the survival probability from Eq.(\ref{30}), Eq.(\ref{31}) and Eq.(\ref{33}) we follow the usual procedure of defining suitable transformations in space and time as in the earlier section. Thus, we define a normalized variable $\bar{X}(t)$ as $\bar{X}(t)=\frac{x(t)}{\sqrt{\langle x^2(t) \rangle}}$ and the correlator in the normalized variable becomes
\begin{eqnarray}
\label{35}
\nonumber
\langle \bar{X}(t_1)\bar{X}(t_2)\rangle &=&\sqrt{\frac{t_2}{t_1}} \quad \textrm{for $0<t<\tau_1$}\\
\nonumber
&=& \sqrt{\frac{\beta L_1^2+2D_2 t_2}{\beta L_1^2+2D_2 t_1}}\quad \textrm{for $\tau_2<t<\tau_3$}\\
\nonumber
&=& \sqrt{\frac{\beta_1 L_1^2+\beta_2 L_2^2+2D_3 t_2}{\beta_1 L_1^2+\beta_2 L_2^2+2D_3 t_1}} \quad \textrm{for $t>\tau_4$} \\
\end{eqnarray}

The time transformation $t\rightarrow T$ for the three different regimes are defined in the following way
\begin{eqnarray}
\label{36}
\nonumber
e^{T} &=& t \quad \textrm{for $0<t<\tau_1$} \\
\nonumber
&=& \beta L_1^2 + 2D_2 t \quad \textrm{for $\tau_2<t<\tau_3$}\\
\nonumber
&=& \beta_1 L_1^2+\beta_2 L_2^2 + 2D_3 t \quad \textrm{for $t>\tau_4$}\\
\end{eqnarray}

and the correlator $\langle \bar{X}(T_1) \bar{X}(T_2) \rangle$ becomes
\begin{eqnarray}
\label{37}
\langle \bar{X}(T_1) \bar{X}(T_2) = e^{-\frac{1}{2}(T_1-T_2)}
\end{eqnarray}
for all the three temporal regimes, the difference being that the time transformations are different in the three regimes. The process is now a Gaussian stationary process.

Since the correlator is exponentially decaying, the survival probability in the transformed time variable is simply
\begin{equation}
\label{38}
 P(T)=e^{-T/2}.
\end{equation}
In real time, using Eq.(\ref{38}), the survival probability becomes
\begin{eqnarray}
\label{39}
\nonumber
 p(t)&\sim&t^{-1/2} \quad \textrm{for $0<t<\tau_1$}\\
 \nonumber
&\sim& \frac{1}{\sqrt{\beta L_1^2+2D_2t}} \quad \textrm{for $\tau_2<t<\tau_3$}\\
&\sim& \frac{1}{\sqrt{\beta_1 L_1^2+\beta_2 L_2^2+2D_3t}} \\
\nonumber
\quad \textrm{for $t>\tau_4$}.
\end{eqnarray}

\begin{figure}
\onefigure[height=7cm,width=8cm]{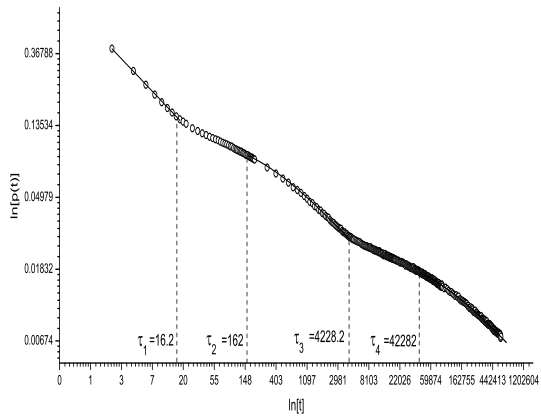}
\caption{Plot of survival probability $p(t)$ vs time in log-log scale for 
$L_1=60$, $L_2=360$, $2D_1=200$, $2D_2=20$ and $2D_3=2$. The crossover time scales and the crossover regime is indicated in the figure. The circular points are actual data from numerical simulation and the solid lines are fit of Eq.(\ref{41}).} 
\end{figure}

\begin{figure}
\includegraphics[height=7cm,width=8cm]{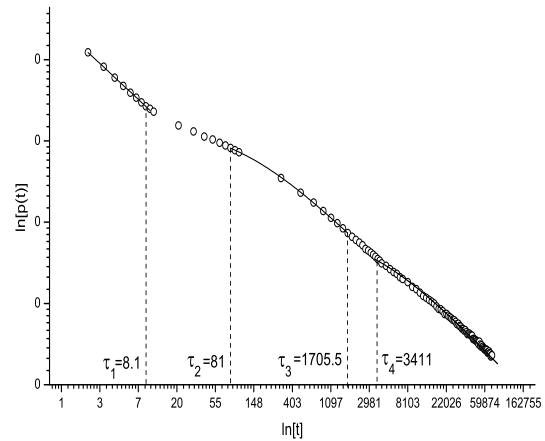}
\caption{Plot of survival probability $p(t)$ vs time in log-log scale for 
$L_1=30$, $L_2=210$, $2D_1=100$, $2D_2=10$ and $2D_3=5$. The crossover time scales and the crossover regime is indicated in the figure. The circular points are actual data from numerical simulation and the solid lines are fit of Eq.(\ref{41}).}
\end{figure}

A plot of the survival probability for two different set of values of $L_1$, $L_2$, $D_1$, $D_2$ and $D_3$ is shown in Fig. 5 and Fig. 6. Configuration averaging of $10^6$ has been done to obtain the numerical results of Fig. 5 and Fig. 6.
Theoretical and numerical values of the time scales are presented in Table D.

\begin{center}
\textbf{Table D}
\end{center}

\begin{tabular}{|c|c|}
\hline
Parameter Values & Time Scales \\
\hline
$L_1=30$ & $\tau_1^{th}=8.1$, $\tau_1^{nu}=8.68$ \\ \cline{2-2}
$L_2=210$& $\tau_2^{th}=81$, $\tau_2^{nu}=80.4393$ \\ \cline{2-2}
$2D_1=100$& $\tau_3^{th}=1705.5$, $\tau_3^{nu}=1702.58$ \\ \cline{2-2}
$2D_2=10$, $2D_3=5$ & $\tau_4^{th}=3411$, $\tau_4^{nu}=3421.27$ \\ \cline{2-2}
\hline
$L_1=60$ & $\tau_1^{th}=16.2$, $\tau_1^{nu}=16.01$ \\ \cline{2-2}
$L_2=360$& $\tau_2^{th}=162$, $\tau_2^{nu}=162.096$ \\ \cline{2-2}
$2D_1=200$& $\tau_3^{th}=4228.2$, $\tau_3^{nu}=4245$ \\ \cline{2-2}
$2D_2=20$, $2D_3=2$ & $\tau_4^{th}=42282$, $\tau_4^{nu}=42329.8$ \\ \cline{2-2}
\hline
\end{tabular}

To conclude, we have investigated the phenomenon of persistence for the case of a random walker in a composite media with two and three homogeneous components. We have presented a very simplified theory to explain the survival probability of a random walker in such inhomogeneous systems. For the two component system, analytical results show that there are two relevant time scales in the problem and this time interval is the crossover regime for the problem. Similarly, for the three component systems there are four relevant time scales and two crossover regimes. The fact that the crossover regimes are not sharp is due to the multiple hoppings that a random walker undergoes near the boundary.

\acknowledgements{D.C acknowledges Council for Scientific and Industrial Research,
Govt. of India for financial support (Grant No.-
9/80(479)/2005-EMR-I). D.C is also  gratefull to J.K.Bhattacharjee for many fruitfull discussions.\\
}

\end{document}